\pdfoutput=1 
\documentclass[prd,twocolumn,amssymb,10pt,aps]{revtex4-1}
\usepackage{amsmath,amssymb,mathrsfs}

\newcommand{\R}{{\mathbb{R}}}
\newcommand{\C}{{\mathbb{C}}}

\newcommand{\T}{{\mathcal{T}}}
\newcommand{\dd}{{{\rm d}}}

\renewcommand{\P}{{\mathcal{P}}}
\newcommand{\PT}{{\mathcal{PT}}}

\renewcommand{\H}{{\mathcal{H}}}
\newcommand{\Dom}{{\rm{Dom}\,}}

\newcommand{\Ran}{{\rm{Ran}}}

\renewcommand{\Re}{\text{\rm Re}\,}
\renewcommand{\Im}{\text{\rm Im}\,}
\newcommand{\etc}{\emph{etc}}
\newcommand{\cf}{\emph{cf.}}

\newcommand{\ie}{{\emph{i.e.}}}
\newcommand{\eg}{{\emph{e.g.}}}
\newcommand{\dist}{\mathrm{dist}}

\newcommand{\eps}{\varepsilon}
\begin{document}
\title{On the metric operator for the imaginary cubic oscillator}
\author{P.~Siegl$^{1,2}$ and D.~Krej\v{c}i\v{r}\'{i}k$^{2}$}
\affiliation{$^{1}$ Group of Mathematical Physics of the University of Lisbon, 
Complexo Interdisciplinar, Av.~Prof.~Gama Pinto~2, 1649-003 Lisboa, 
Portugal. 
\\
$^{2}$ Department of Theoretical Physics, Nuclear Physics Institute ASCR, 
25068 \v{R}e\v{z},  Czech Republic.
}

\begin{abstract}
We show that the eigenvectors of the $\PT$-symmetric imaginary cubic oscillator 
are complete, but do not form a Riesz basis. This results in the existence
of a bounded metric operator having intrinsic singularity reflected
in the inevitable unboundedness of the inverse. Moreover, the existence 
of non-trivial pseudospectrum is observed.
In other words, there is no quantum-mechanical Hamiltonian 
associated with it via bounded and boundedly invertible 
similarity transformations.
These results open new directions in physical interpretation 
of $\PT$-symmetric models with intrinsically singular metric, 
since their properties are essentially different
with respect to self-adjoint Hamiltonians, 
for instance, due to spectral instabilities.
\begin{center}
\fbox{Published in: Physical Review D 86 (2012), 121702(R).}
\end{center}
\end{abstract}

\maketitle


\section{Introduction}

At the turn of the millennium, 
Bender \emph{et al.}\ came up with the idea  
to extend quantum mechanics by considering Hamiltonians
that are invariant under a space-time reflection~$\mathcal{PT}$
rather than being Hermitian
\cite{Bender-1998-80,Bender-2002-89}.  
The development of the so-called 
\emph{$\mathcal{PT}$-symmetric quantum mechanics} 
was in fact initiated in these papers
by considering a prominent Hamiltonian
\begin{equation}\label{H.def}
  H = -\frac{\mathrm{d}^2}{\mathrm{d}x^2} + ix^3
  .
\end{equation}
While this operator is manifestly non-Hermitian,
it is invariant under a simultaneous 
space reflection~$\mathcal{P}$ ($x \mapsto -x$) 
and time reversal~$\mathcal{T}$ (complex conjugation).
Moreover, numerical studies suggested that the spectrum of~$H$
is \emph{real}, 
which was later proved in \cite{Dorey-2001-34,*Shin-2002-229}.
The Hamiltonian~\eqref{H.def} can be considered as a prototype
of many other examples of $\mathcal{PT}$-symmetric Hamiltonians
that have been so far studied in a still growing literature
(see 
\cite{Bender-2007-70,*Mostafazadeh-2010-7} and references therein).
$\PT$-symmetric models found 
applications in various domains of physics
-- namely in optics 
\cite{Regensburger-2012-488,*Longhi-2009-103,*Guo-2009-103,*Klaiman-2008-101,*Ruter-2010-6}, 
solid state \cite{Bendix-2009-103}, Bose-Einstein condensates \cite{Graefe-2012},
LRC circuits \cite{Schindler-2011-84,*Lin-2012-85}, 
superconductivity \cite{Rubinstein-2007-99,*Rubinstein-2010-195}, 
electromagnetism \cite{Ruschhaupt-2005-38,*Mostafazadeh-2009-102}, 
and reflectionless scattering \cite{Hernandez-Coronado-2011-375}.

It is commonly accepted that 
a \emph{quantum-mechanical} interpretation of $\mathcal{PT}$-symmetry
must be implemented through a similarity transformation $\Omega$, 
\ie
\begin{equation}\label{sim.transf}
h:=\Omega H \Omega^{-1},
\end{equation}
where $h$ is a self-adjoint operator, \ie~$h=h^\dagger$. 
This intertwining relation is closely related 
to the \emph{quasi-Hermiticity}~\cite{Scholtz-1992-213, Dieudonne-1961}
\begin{equation}\label{Theta.def}
\Theta H = H^{\dagger} \Theta,
\end{equation}
where $\Theta$ is a positive operator often called \emph{metric operator} 
(its special variant $\P \mathcal{C}$ 
was suggested in Refs.~\cite{Bender-2002-89,Bender-2004-70}).
Hamiltonian~$H$ with property~\eqref{Theta.def} is called
quasi-Hermitian because it is actually Hermitian with respect to the modified inner product 
$\langle \cdot , \Theta \cdot \rangle$. 
The relation between $\Omega$ and $\Theta$ 
is the decomposition of a positive operator $\Theta = \Omega^{\dagger}\Omega$.
The essential idea is that a non-Hermitian~$H$ can be viewed as an alternative
representation of a Hermitian operator~$h$. 

The advantage of the above described representation~\eqref{sim.transf} 
stems from the observation that the Hermitian counterpart~$h$ 
for a differential albeit non-Hermitian operator~$H$
has typically a non-local and very complicated structure. 
This was demonstrated for a class of operators
with non-Hermitian (not necessarily $\PT$-symmetric) point interactions 
in \cite{Krejcirik-2012,*Krejcirik-2006-39,*Krejcirik-2008-41a},
where, in addition, explicit formulae 
for the similarity transformation~$\Omega$, 
metric operator~$\Theta$, $\mathcal{C}$ operator, 
and similar self-adjoint operator~$h$
were presented in a closed form. 
Nevertheless, the non-Hermiticity 
and non-locality are not always equivalent in the described sense 
\cite{Albeverio-2005-38,Siegl-2008-41,*Kuzhel-2011-379}. 

Partly motivated by the relevance of the cubic interaction
in quantum field theory, 
the problem of similarity of the Hamiltonian~\eqref{H.def}
to a self-adjoint operator was investigated in several works
\cite{Bender-2004-70,*Bender-2005-71a,Mostafazadeh-2006-39a}.
However, due to the complexity of the task,
the approach used in these papers was necessarily formal,
based on developing the metric into an infinite series
composed of unbounded operators. 
There has been no proof of 
the quasi-Hermiticity of the imaginary cubic oscillator so far.
The objective of the present note is 
to establish the following intrinsic facts 
about the metric of~\eqref{H.def}:
\begin{enumerate}
\item
\emph{There exists a bounded metric.} 
That is, operator~\eqref{H.def}
is quasi-Hermitian in the sense of~\eqref{Theta.def}
with bounded~$\Theta$.
\item
\emph{No bounded metric with bounded inverse exists.}
That is, any metric operator for~\eqref{H.def}
necessarily possesses an inevitable singularity.
\end{enumerate}

We have chosen the prominent Hamiltonian~\eqref{H.def}
to prove the negative result~2 just because 
the $ix^3$ potential is considered as the \emph{fons et origo}
of $\mathcal{PT}$-symmetric quantum mechanics 
\cite{Bender-1998-80,Bender-2002-89}.
However, the absence of bounded or boundedly invertible metric
is by far not restricted to the Hamiltonian~\eqref{H.def} only.
For instance, the method of the present note
also applies to an equally extensively studied
$x^2+i x^3$ potential and many others, 
see Eq.~\eqref{Davies.potential} and the surrounding text.

Our results have important consequences for the physical 
interpretation of the $\PT$-symmetric Hamiltonians.
If the metric happens to be singular 
(\ie~unbounded, not invertible or unboundedly invertible),
the quantum-mechanical interpretation 
using the similarity transformation is lost.
Indeed, the eigenvectors, despite possibly being complete,
do not form a ``good'' basis, \ie~an unconditional (Riesz) basis. 
The spectrum of such highly non-self-adjoint operators 
does not contain sufficient information
about the system and in addition to the reality 
and (algebraic) simplicity of the spectrum, 
more involved spectral-theoretic properties 
(such as basicity, pseudospectrum, \emph{etc}.) 
must be taken into account. 

Our result about the singularity of any metric 
may seem negative at the first glance.
However, we believe that in the same way as 
the exceptional points represent
one of the most interesting configurations, 
where important physical phenomena arise, 
the established intrinsic singularities in the metric operator 
are precisely the point
where new developments of the physics of $\PT$-symmetric models may originate.

This paper is organized as follows.
In Section~\ref{Sec.vs} we emphasize some aspects 
of unbounded operators and defects of quasi-Hermiticity
based on singular metrics.
In Section~\ref{Sec.ix3} we recall known facts about
the imaginary cubic oscillator and perform our proofs
of the new properties regarding the metric operator. 
Finally, in Section~\ref{Sec.end} we refer to some open problems
and comment on possible extensions of our results.

\section{Infinite-dimensional subtleties}\label{Sec.vs}
%
While the concepts of similarity to a self-adjoint operator 
and quasi-Hermiticity
work smoothly if the dimension of 
the underlying Hilbert space is finite, 
\ie~for matrices, 
essential difficulties may appear in the infinite-dimensional spaces.
The reason is obviously in the unboundedness of operators, 
which unavoidably restricts their domains of definitions
to a non-trivial subset of the Hilbert space. 
Therefore,
the sense in which equalities \eqref{sim.transf} and~\eqref{Theta.def}
hold must be carefully explained. We focus on the metric operator further, 
nonetheless, the similarity transformation 
may be discussed along the same lines.

Relation~\eqref{Theta.def} is an operator equality and as such 
it requires that the operator domains $\Dom(\Theta H)$
and $\Dom(H^\dagger \Theta)$ are equal in addition to
the validity of the corresponding vector identity
$
  \Theta H \psi = H^{\dagger} \Theta \psi
$
for every $\psi \in \Dom(\Theta H)\cap\Dom(H^\dagger \Theta)$.
Problems arise if the involved operators are unbounded,
since one of the operator domains of the products 
or their intersection might be reduced to a single element $0$. 
To avoid such pathological situations, 
it is usually assumed that the metric operator~$\Theta$ is bounded.
Then the above requirements reduce 
to the mapping property $\Theta [\Dom(H^{\dagger})] \subset \Dom(H)$
and the quasi-Hermitian identity should hold for every $\psi \in \Dom(H)$.

If, in addition to the boundedness, $\Theta$ is boundedly invertible,
then some fundamental 
and extremely useful properties of 
self-adjoint operators are valid for~$H$ as well:
real spectrum, spectral decomposition,
spectral stability with respect to perturbations, 
unitary evolution (in a topologically equivalent Hilbert space), \etc. 
However, if the metric becomes singular, 
none of the mentioned properties is guaranteed 
by the validity of~\eqref{Theta.def}. 
As a matter of fact, as we demonstrate in this paper, 
the imaginary cubic oscillator and many other $\PT$-symmetric Hamiltonians, 
despite possessing real spectra, 
exhibit pathological features with respect to self-adjoint behaviour,
due to the intrinsic singularities of the metric 
(and therefore also in $\mathcal{C}$-operators 
and similarity transformations).
Let us demonstrate the defects of theories with singular metrics
in the following subsections.

\subsection{Spectrum}
Let~$\H$ be a complex Hilbert space.
The \emph{spectrum} is meaningfully defined only for closed operators,
\ie~those operators $H$ for which
the elements $\{\psi,H\psi\}$ with $\psi \in \Dom(H)$ form 
a closed linear subspace of $\H\times\H$.
If~$\H$ were finite-dimensional, then the spectrum of~$H$, $\sigma(H)$,
would be exhausted by eigenvalues, \ie~those complex numbers~$\lambda$  
for which $H-\lambda$ is not injective.
In general, however, there are additional parts of spectra composed  
by those~$\lambda$ which are not eigenvalues 
but $H-\lambda:\Dom(H)\to\H$ is not bijective:
depending on whether the range $\Ran(H-\lambda)$ 
is dense in~$\H$ or not, one speaks about the \emph{continuous}
or \emph{residual} spectrum, respectively.
In other words, the complement of the spectrum of~$H$,
called the \emph{resolvent set} of~$H$, $\rho(H)$, 
is composed of all the complex numbers~$z$ 
for which the resolvent operator $(H-z)^{-1}:\H\to\H$ 
exists and is bounded.  

It is an important property of self-adjoint operators
that their (total) spectrum is always non-empty,
real and that the residual spectrum is empty. 
For non-self-adjoint operators, however,
the spectrum can be empty or cover the whole complex plane,
see \eg~\cite{Albeverio-2005-38,Almog-2008-40}.

Let us demonstrate how singular metrics lead to 
pathological situations as regards spectral properties.
Let~$H$ be an operator with purely discrete spectrum 
(\ie~just isolated eigenvalues with finite multiplicities)
and assume that the similarity relation~\eqref{sim.transf} 
holds with unbounded~$\Omega^{-1}$.
Then $\Ran(h-\lambda) \subset \Ran(\Omega) \not= \H$
for every $\lambda \in \C$.
Consequently, the whole complex plane
except for the set of eigenvalues of~$H$
belongs to the continuous spectrum of~$h$.
Summing up, \emph{the continuous spectrum 
is not preserved by unbounded similarity transformations}.
It is a striking phenomenon since 
the continuous part of spectrum contains physical energies 
corresponding to scattering/propagating states.

A similar argument shows that unbounded~$\Theta$ 
satisfying~\eqref{Theta.def} with $\Dom(\Theta)\supset\Dom(H)$ 
and $\Theta [\Dom(H)] \subset \Dom(H^\dagger)$
cannot exist for closed operators~$H$ with a physically reasonable 
property $\sigma(H)\not=\C$. 
In this way, one can also show that the $\mathcal{C}$-operator 
of~\cite{Kuzhel-2009-190} for \eqref{H.def} cannot exist.

\subsection{Eigenbasis}
Eigenfunctions of self-adjoint operators 
corresponding to different eigenvalues
are mutually orthogonal.
Furthermore, the set of all eigenfunctions $\{\psi_n\}_{n=1}^\infty$ 
of a self-adjoint operator with purely discrete spectrum can be normalized
in such a way that it forms a \emph{complete orthonormal family}
in the Hilbert space~$\H$.
Recall that the completeness means that the orthogonal 
complement in~$\H$ of the linear span of the family
consists only of the zero function only.
A necessary and sufficient condition for completeness
of an orthonormal family $\{\psi_n\}_{n=1}^\infty$
is the validity of the Parseval equality
\begin{equation}\label{Parseval}
\forall \psi \in \H, \qquad  \sum_{n=1}^\infty |\langle\psi_n,\psi\rangle|^2 = \|\psi\|^2.
\end{equation}
In this case we also have the unique expansion
\begin{equation}\label{basis}
\forall \psi \in \H, \qquad  \psi = \sum_{n=1}^\infty c_n \psi_n.
\end{equation}
That is, $\{\psi_n\}_{n=1}^\infty$ is a \emph{basis} in~$\H$.

Eigenfunctions of non-Hermitian operators are typically not orthogonal.
Even worse, they may not form a basis or even not a complete family.
In this respect, it is absolutely essential to stress 
that the completeness of a non-orthonormal family $\{\psi_n\}_{n=1}^\infty$
does not imply that any $\psi\in\H$ admits a unique expansion~\eqref{basis};
see \eg~\cite{Davies-2007} for further details.

The notion of ``eigenbasis'' is so important in quantum mechanics 
that one needs to have a replacement for~\eqref{Parseval}
in the case of eigenfunctions of non-Hermitian operators. 
This is provided by the notion of \emph{Riesz basis}
\begin{equation}\label{Riesz}
\forall \psi \in \H, \qquad
  C^{-1} \|\psi\|^2
  \leq \sum_{n=1}^\infty |\langle\psi_n,\psi\rangle|^2 \leq 
  C \|\psi\|^2
\end{equation}
with a positive constant $C$ independent of $\psi$.
Eigenfunctions of an operator~$H$ with purely discrete spectrum 
form a Riesz basis if, and only if, $H$~is quasi-Hermitian~\eqref{Theta.def} 
with bounded and boundedly invertible metric~$\Theta$. 

As in the case of spectrum, 
\emph{Riesz-basicity property is not preserved by unbounded transformations}.
As a matter of fact, it is the objective of the present paper to show
that the eigenfunctions of~\eqref{H.def} do not form a Riesz basis,
so that the metric~$\Theta$ is necessarily singular.
Any claim of the type 
``\eqref{H.def} is similar to a self-adjoint operator''
is thus necessarily of doubtful usefulness for physics,
since~$H$ and~$h$ appearing in~\eqref{sim.transf}
would have very different basicity properties. 

\subsection{Pseudospectrum}
The notion of pseudospectra arose as a result of the realization 
that several pathological properties of highly non-Hermitian 
operators were closely related. 
We refer to by now classical monographs 
by Trefethen and Embree~\cite{Trefethen-2005}
and Davies~\cite{Davies-2007} for more information
on the subject, physical and numerical applications, 
and many references.

Given a positive number~$\eps$,
we define the \emph{pseudospectrum} of~$H$ by
\begin{equation}\label{pseudo}
  \sigma_\eps(H) := 
  \big\{
  z \in \C \ \big| \ 
  \|(H-z)^{-1}\| > \eps^{-1} 
  \big\}
  \,,
\end{equation}
with the convention that $\|(H-z)^{-1}\|=\infty$ for $z \in \sigma(H)$.
The pseudospectrum always contains an $\eps$-neighbourhood 
of the spectrum:
\begin{equation}
  \big\{z\in\C\ \big| \ \dist\big(z,\sigma(H)\big) < \eps \big\}
  \subset \sigma_\eps(H) 
  \,.
\end{equation}
Since equality holds here if~$H$ is self-adjoint 
(or more generally normal), it follows that 
the notion of pseudospectra becomes trivial for such operators.
On the other hand, if~$H$ is ``highly non-self-adjoint'',
the pseudospectrum $\sigma_\eps(H)$ is typically ``much larger''
than the $\eps$-neighbourhood of the spectrum. 

For non-Hermitian operators the pseudospectra
are much more reliable objects than the spectrum itself. 
Probably the strongest support for this claim is due to
phenomenon of \emph{spectral instability}:
very small perturbations may drastically change 
the spectrum of a non-Hermitian operator.
For instance, new complex eigenvalues can appear 
very far from the original ones. 
On the other hand, perturbations whose norm is less than~$\eps$
still lie inside~$\sigma_\eps(H)$.   
These effects were extensively studied in numerics, 
hydrodynamics, optics, \etc.\
(see~\cite{Trefethen-2005} and references therein).

Of course, such pathological situations do not occur
for self-adjoint operators whose spectrum is changed at most
by the norm of the perturbation.
It is also impossible for operators similar to self-adjoint operators
by bounded and boundedly invertible similarity transformations.
On the other hand, \emph{the pseudospectrum is not preserved
by unbounded transformations}
(we refer to~\cite{Zworski-2002-229} for a warning discussion of 
the shifted harmonic oscillator in this context).
The pseudospectrum thus represents a useful test whether
a given non-Hermitian operator can be similar to a self-adjoint one
via a physically reasonable transformation.
In this paper we show that the pseudospectrum of~\eqref{H.def}
is highly non-trivial.

\subsection{Singular metric?}
\label{subsec.sing.m}

The observations made in previous subsections constitute 
a strong support for our belief that the singular metrics
are not relevant objects for physical interpretation 
of non-Hermitian Hamiltonians, since they yield
only singular similarity transformations.
However, putting it differently,
singular metrics necessarily lead to fundamentally new physics,
since the transformed operators exhibit completely different properties.

In this context we feel necessary 
to mention that there exists a recent attempt 
of Mostafazadeh~\cite{Mostafazadeh-2012}, 
reproducing equivalently the original idea of 
Kretschmer and Szymanowski~\cite{Kretschmer-2004-325}, 
to include singular metric operators into the notion of quasi-Hermiticity.
It involves a construction of a self-adjoint operator
to which the original non-Hermitian operator with purely discrete
real spectrum is similar ``at any cost''.
Analogous ideas for unbounded $\mathcal{C}$-operators 
can be found in \cite{Bender-2012-45}.
However, any such strategy has important drawbacks that cannot
be avoided. The problem with singular metric 
is mentioned already in \cite{Dieudonne-1961}, 
where an example of operator possessing bounded metric operator 
without bounded inverse and 
having \emph{non-real spectrum} was constructed.
As a corollary, Diedonn\'e states: 
\emph{``in spite of the quasi-Hermiticity (without bounded inverse of $\Theta$), 
there is for instance no hope of building functional calculus that would follow more or less the same pattern 
as the functional calculus of self-adjoint operators''}.

The drawbacks consist in that the aforementioned 
non-self-adjoint pathologies of~$H$ are completely ignored 
when analysing the ``similar'' self-adjoint operator~$h$ instead.
This can be illustrated already for two-by-two matrices: 
a Jordan-block matrix~$H$ and a diagonal matrix~$h$ 
with the same real eigenvalues.
Although the matrices possess the same (real) spectrum,
their respective properties are very different, 
particularly the basicity properties of eigenvectors 
and spectral stability with respect to small perturbations. 
But the construction of \cite{Kretschmer-2004-325,Mostafazadeh-2012}, 
when used in finite dimension, simply
means that the authors disregard the Jordan-block structure 
of the non-Hermitian matrix~$H$ and associate to it 
just the diagonal matrix~$h$ with same eigenvalues. 
The metric operator and ``similarity transformation'' 
are non-invertible in this case.
However, equality \eqref{Theta.def} 
and a weaker variant of \eqref{sim.transf}, \ie~$\Omega H= h \Omega$, do hold. 
Stating that~$h$ should in any reasonable sense represent~$H$ 
is obviously very doubtful,
since, for instance, all the physics of exceptional points 
would be omitted.

In infinite-dimensional spaces the situation is even more complex,
since another possibility of singularity of metric exists, namely the
unboundedness of the inverse. Although this may seem to be 
a minor issue or only a technical problem of infinite dimension, 
such an interpretation is very misleading.
The pathological properties of non-self-adjoint~$H$ 
with only unboundedly invertible metric may be
much more serious than existence of finite-dimensional Jordan blocks, 
\ie~usual exceptional points.
In the latter case, although the metric cannot be invertible,
the eigenvectors together with generalized ones may 
form a Riesz basis. In other words, except a finite-dimensional
subspace, $H$~is similar to a self-adjoint operator. 
Therefore a version of the spectral decomposition 
(generalized Jordan form) may be available and the spectrum of~$H$ 
may be stable with respect to small perturbations. 
This is not the case of the imaginary cubic oscillator, 
where there is no Riesz basis of eigenvectors and
no spectral stability: complex eigenvalues may appear very 
far from the unperturbed real ones despite the norm
of the perturbation is arbitrarily small.

\section{Imaginary cubic oscillator}\label{Sec.ix3}
%
Let us begin by properly introducing the Hamiltonian~\eqref{H.def}
as a closed realization in the Hilbert space $L^2(\R)$.
We consider the \emph{maximal realization}
of the differential expression~\eqref{H.def}  
by taking for the operator domain of~$H$
the maximal domain
\begin{equation}
  \Dom(H) := \left\{
  \psi \in L^2(\R) \, | \,
  -\psi''+ix^3\psi \in L^2(\R)
  \right\}
  .
\end{equation}
By an approach of~\cite[Sec.~VII.2]{EE}, 
based on a distributional Kato's inequality,
it follows that such a defined operator~$H$ is \emph{m-accretive}
and that it coincides with the closure of~\eqref{H.def} 
initially defined on infinitely smooth functions of compact support.
(The difficulties with the existence of different closed extensions, 
\cf~\cite{Azizov-2010-43,*Azizov-2012-53}, 
do not arise here since, $\Re V$ is trivially bounded from below.)

Now it can be rigorously verified that~$H$ is 
\emph{$\mathcal{PT}$-symmetric}, \ie\ $[H,\mathcal{PT}]=0$, 
where the commutator should be interpreted as 
$\PT H \psi = H \PT \psi$ for all $\psi \in \Dom(H)$,
with $(\mathcal{P}\psi)(x):=\psi(-x)$ 
and $(\mathcal{T}\psi)(x):=\overline{\psi(x)}$.
Moreover, since the adjoint~$H^{\dagger}$ of~$H$ is simply obtained
by taking~$-i$ instead of~$i$ in the definition 
of the operator (including the operator domain),
it can be also verified that~$H$ is \emph{$\mathcal{P}$-self-adjoint},
$H^{\dagger}=\mathcal{P}H\mathcal{P}$,
and \emph{$\mathcal{T}$-self-adjoint},
$H^{\dagger}=\mathcal{T}H\mathcal{T}$.
The latter is a particularly useful property 
for non-self-adjoint operators
since it implies that the residual spectrum of~$H$
is empty \cite{Borisov-2008-62}.

As an immediate consequence of the fact that~$H$ is m-accretive,
we know that the spectrum of~$H$ is located 
in the right complex half-plane.
Furthermore, it has been shown in 
\cite{Dorey-2001-34,Shin-2002-229}
that \emph{all eigenvalues of~$H$ are real} and simple
(in the sense of geometric multiplicity).
The algebraic simplicity has been established in \cite{Tai-2005-38,*Tai-2006-223}.
The fact that the \emph{spectrum of~$H$ is purely discrete}
follows from the compactness of its resolvent. 
The latter can be deduced from the identity
\begin{equation}\label{compact}
  \Dom(H) = \left\{
  \psi \in H^2(\R) \, | \,
  x^3\psi \in L^2(\R)
  \right\}
\end{equation}
established in~\cite{Caliceti-1980-75}
and the compact embedding of this set into~$L^2(\R)$.
Furthermore, the authors of~\cite{Caliceti-1980-75}
show that the resolvent of~$H$ is a Hilbert-Schmidt operator.
The key ingredient in the proof is the explicit knowledge 
of the resolvent kernel of~$H^{-1}$ that can be written 
in terms of Hankel functions with known asymptotics. 
A deeper analysis of the resolvent of~$H$ 
reveals that it actually belongs 
to the trace class~\cite{Mezincescu-2001-34};
alternatively, one can use a general result
of~\cite{Robert-1978-3}.

\subsection{Completeness of eigenfunctions}

Let us show that the eigenfunctions of~$H$ 
form a complete set in $L^2(\R)$.
Recall that the completeness of~$\{\psi_n\}_{n=1}^\infty$ 
means that the span of $\psi_n$ is dense in $L^2(\R)$, 
or equivalently $\big({\rm span} \{\psi_n\}_{n=1}^\infty)^{\perp} = \{0\}$. 
Nevertheless, we stress that the result on completeness does not imply
that any~$\psi$ admits the unique expansion~\eqref{basis}.

The m-accretivity of~$H$ implies 
$\Re \langle \psi, H \psi \rangle \geq 0$
for all $\psi \in \Dom(H)$.
Consequently, $-iH$~is dissipative, 
\ie~$\Im \langle \psi, H \psi \rangle \leq 0$
for all $\psi \in \Dom(H)$.
It is then easy to check that 
the imaginary part of the resolvent of 
$-i H$ at $\xi<0$ is non-negative, \ie,
\begin{equation}
\frac{1}{2i}\left((-i H-\xi)^{-1} - (i H^{\dagger}-\xi)^{-1}\right) \geq 0
\end{equation}
in the sense of forms.
Since the resolvent is trace class, 
it is enough to apply the completeness theorem 
\cite[Thm.VII.8.1]{Gohberg-1990} to the operator $(-i H-\xi)^{-1}$.

More specifically, it follows 
by this result that~$H$ has a complete system of eigenvectors
and generalized eigenvectors. The latter, however, 
do not appear in our situation since all the eigenvalues
are algebraically simple (see above).

\subsection{Existence of a bounded metric}

Already at this stage, we can show that there exists a bounded metric for $H$.
We would like to emphasize that this follows actually in general from the reality and 
simplicity of eigenvalues and completeness of eigenfunctions for $H$. We remark that 
$H^{\dagger}$ shares these properties due to the simplicity of eigenvalues 
and~$\T$- or~$\P$-self-adjointness of $H$.  

In detail, 
let~$H$ be a densely defined 
and closed operator such that 
$\rho(H) \cap \rho(H^{\dagger}) \cap \R \neq \emptyset$ and let
$z_0$ be a number from this intersection.
Then the existence of a bounded positive~$\Theta$ 
satisfying~\eqref{Theta.def} 
is equivalent to the fact that the resolvent of~$H$ 
satisfies \eqref{Theta.def}, \ie,
\begin{equation}
\Theta (H-z_0)^{-1} = (H^{\dagger}-z_0)^{-1} \Theta.
\end{equation}
Thus we can transfer the problem of finding 
the metric for an unbounded~$H$ to the same problem 
for its bounded resolvent. 
Using~\cite[Prop.3]{Dieudonne-1961}, 
which is in fact the construction of a bounded metric using the well-known formula 
\begin{equation}
\Theta := \sum_{n=1}^\infty c_n \phi_n \langle \phi_n, \cdot \rangle
\end{equation}
with~$\phi_n$ being the eigenfunctions of~$H^\dagger$ 
and~$c_n>0$ tending to zero sufficiently fast, yields the following: 
If all the eigenvalues of~$H^{\dagger}$ are real 
and the associated eigenfunctions~$\phi_n$ are complete, 
then a bounded metric for $(H-z_0)^{-1}$, and therefore for $H$, exists.

In our situation, we know that all the eigenvalues of~\eqref{H.def} as well as its adjoint
are simple and real, for~$z_0$ we can take any negative number
due to m-accretivity of~$H$ and $H^{\dagger}$, 
and we have shown that the eigenfunctions of~$H$ and therefore also $H^{\dagger}$ are complete. 
Hence the existence of a bounded~$\Theta$ follows.

\subsection{Singularity of any metric}
\label{subsec.res.b}

After the two preceding positive results,
we show now that any metric for the imaginary cubic oscillator is singular, 
\ie~either unbounded or unboundedly invertible.
We proceed by contradiction. 
Let there exist a bounded positive operator~$\Theta$ 
with bounded inverse satisfying~\eqref{Theta.def}.
Then the following norm estimate for the resolvent holds:
\begin{equation}\label{rbound}
  \big\| (H-z)^{-1}\big\| \leq \frac{C}{|\Im z|}
\end{equation}
for every $z \in \C$ such that $\Im z \not= 0$, 
where~$C$ is a positive constant bounded 
by $\big\|\sqrt{\Theta}\big\|\big\|\sqrt{\Theta^{-1}}\big\|$. 
By establishing a lower bound to the resolvent appearing in~\eqref{rbound}, 
we show that the inequality~\eqref{rbound} cannot hold. 
The lower bound follows by a direct construction 
of a continuous family of approximate eigenstates
of complex energies far from the spectrum
due to Davies \cite{Davies-1999-200}. 

Using the strategy in \cite[Thm.~2]{Davies-1999-200}, 
we consider $\|(H-\sigma z)^{-1}\|$ with $\sigma>0$ large 
and $0 < \arg z < \pi/2$. 
By a simple scaling argument in~$x$,
the problem can be transferred into a semi-classical one, 
namely $\|(H-\sigma z)^{-1}\| = \sigma^{-1} \|(H_h- z)^{-1}\|$,
where 
\begin{equation}
H_h:= - h^2 \frac{\dd^2}{\dd x^2} + i x^3, 
\end{equation}
with $h:=\sigma^{-5/6}$. 
In order to apply \cite[Thm.~1]{Davies-1999-200}, 
we have to verify that $\Im V'(a) \neq 0$, 
where $V(x):=ix^3$ 
and~$a$ is obtained from the relation 
$z=\eta^2 + i a^3$ with $\eta \in \R\setminus\{0\}$. 
However, this can be easily checked for $\Im V'(a) = 3 a^2$ 
and $a\neq 0$ since $\Im z \neq 0$ by assumption. 
It then follows from \cite[Thm.~1]{Davies-1999-200} 
that the norm of the resolvent of $H_h$ diverges faster 
than any power of $h^{-1}$ as $h\rightarrow 0$. 
More specifically, there exists positive~$h_0$ 
and for each $n>0$ a positive constant~$c_n$ such that
if $h\in(0,h_0)$ then
\begin{equation}\label{lbound}
  \big\| (H_h-z)^{-1}\big\| \geq \frac{c_n}{h^n}
  \,.
\end{equation}
The relation between~$H$ and~$H_h$ provides an analogous claim 
for $\|(H-\sigma z)^{-1}\|$ 
and therefore the resolvent bound~\eqref{rbound} 
when combined with~\eqref{lbound} 
cannot hold if~$n$ is chosen sufficiently large (namely, $n>6/5$). 

\section{Concluding remarks}\label{Sec.end}
%
Although the imaginary cubic oscillator~\eqref{H.def} is
$\mathcal{PT}$-symmetric with purely real and discrete spectrum,
it cannot be similar (via a bounded and boundedly invertible transformation)
to any self-adjoint operator or, equivalently, the eigenfunctions of~$H$ cannot 
form the Riesz basis. 
We remark that the question whether eigenvectors of~$H$ form a basis remains open.

We established the existence of a bounded metric, 
which is in fact equivalent to the completeness of
eigenfunctions that we proved 
and the reality and simplicity of eigenvalues.
However, the singular nature of any metric is inevitable. 
The latter was established by semiclassical tools,
namely the pseudomode construction due to~\cite{Davies-1999-200}. 
The method of proof implies that~\eqref{H.def} 
possesses a very non-trivial pseudospectrum
and regions of strong spectral instabilities,
\cf~\eqref{pseudo} and \eqref{lbound}. 
In the language of exceptional points, 
the imaginary cubic oscillator possesses an ``intrinsic exceptional point'' 
that is much stronger than any exceptional point associated 
with finite Jordan blocks, 
\cf~subsections \ref{subsec.sing.m}, \ref{subsec.res.b}. 

The method of this paper, namely the disproval of quasi-Hermiticity
with bounded and boundedly invertible metric
based on the localized semiclassical pseudomodes,
does not restrict to the particular Hamiltonian~\eqref{H.def}.  
It also applies to the already mentioned $x^2+i x^3$ potential, 
and to many others. 
As a large class of admissible operators
let us mention for instance the Schr\"odinger operators considered 
by Davies~\cite{Davies-1999-200}:
\begin{equation}\label{Davies.potential}
  - \frac{\mathrm{d}^2}{\mathrm{d}x^2} 
  + \sum_{m=1}^{2 n} c_m x^{m}
  \,,
\end{equation}
where the constant~$c_{2n}$ has positive real and imaginary parts;
then the corresponding closed realization is an m-sectorial operator.
Later, the results of~\cite{Davies-1999-200} 
were substantially generalized to higher dimensions 
and pseudodifferential operators in 
\cite{Zworski-2001-129,Dencker-2004-57}.

\section*{Acknowledgements}
We are grateful to K.~C.~Shin for informing us about the results 
on algebraic multiplicities of the imaginary cubic oscillator 
and to D.~Robert for telling us about 
his general result on Schatten-class properties
of non-self-adjoint Schr\"odinger operators.
We thank J. Mare\v s for his support and help during
the preparation of this paper.

This work has been supported by a grant within the scope of FCT's project PTDC/\ MAT/\
101007/2008 and partially supported by FCT's projects PTDC/\ MAT/\
101007/2008 and PEst-OE/MAT/UI0208/2011 
and by the GACR grant No.\ P203/11/0701. 

\bibliographystyle{apsrev4-1}
\bibliography{references}

\end{document}